%% file: paper.tex
\documentclass[letterpaper, 10 pt, conference]{ieeeconf}
\IEEEoverridecommandlockouts                              
\usepackage{bmpsize}
\usepackage{xspace}
\usepackage[hyphens]{url}
\usepackage{amsmath}
\usepackage{listings}
\usepackage{balance}
\usepackage{textcomp}
\usepackage{epsfig,graphics,epigraph,graphicx,subfigure, color,xcolor}
\newcommand{\name}{AVG$\textsc{uard}$\xspace}

\definecolor{dkgreen}{rgb}{0,0.7,0}
\title{\LARGE \bf
Towards Secure and Safe Appified Automated Vehicles
}

\author{Yunhan Jack Jia$^{1}$, Ding Zhao$^{2}$, Qi Alfred Chen$^{1}$, Z.~Morley Mao$^{1}$
\thanks{* The first two authors, Y. Jia and D. Zhao, have equal contributions to this research.}
\thanks{$^{1}$Y. Jia, Q. Chen, and Z. Mao are with the Department of Electrical Engineering and Computer Science, Ann Arbor, MI, 48109, USA ({\tt\small jackjia@umich.edu, \tt\small alfchen@umich.edu, \tt\small zmao@umich.edu})}%
\thanks{$^{2}$Ding Zhao is with the University of Michigan Transportation Research Institute, Ann Arbor, MI 48109,USA  (corresponding author: {\tt\small zhaoding@umich.edu})}%
}

\begin{document}


\maketitle


\IEEEpeerreviewmaketitle

\setcounter{page}{1}
\input{secs/abstract}
\input{secs/introduction}



\input{secs/overview}


%
%
%
\input{secs/standard}

\input{secs/static}

\input{secs/dynamic}
\input{secs/guardian}
\input{secs/discussion}

\input{secs/conclusion}


\bibliographystyle{abbrv}
\balance
\bibliography{reference}
\end{document}

%% file: secs/abstract.tex
\begin{abstract}
The advancement in Autonomous Vehicles (AVs) has created an enormous market for the development of self-driving functionalities, raising the question of how it will transform the traditional vehicle development process. One adventurous proposal is to open the AV platform to third-party developers, so that AV functionalities can be developed in a crowd-sourcing way, which could provide tangible benefits to both automakers and end users. Some pioneering companies in the automotive industry have made the move to open the platform so that developers are allowed to test their code on the road. Such openness, however, brings serious security and safety issues by allowing untrusted code to run on the vehicle. In this paper, we introduce the concept of an \emph{Appified} AV platform that opens the development framework to third-party developers. To further address the safety challenges, we propose an enhanced appified AV design schema called \name, which focuses primarily on mitigating the threats brought about by untrusted code, leveraging theory in the vehicle evaluation field, and conducting program analysis techniques in the cybersecurity area. Our study provides guidelines and suggested practice for the future design of open AV platforms.   
\end{abstract}

%% file: secs/introduction.tex
\section{Introduction}

\emph{Appified} platforms, where software applications (apps) are developed in a crowd-sourcing manner by third party developers and distributed through the app market, have achieved astonishing success in the IT field in the last decade due to the benefits accrued from its open nature. Taking the smartphone industry as an example, we can see that it took only four years for the appified platforms -- iOS and Android, which provide an unprecedented rich set of functionalities to users through their app stores -- to win 91.1\% of the market share since their birth in 2008~\cite{idc_market_Share}. Recent years have seen the \emph{appification} of many other software platforms such as the smart home~\cite{smartthings}, network switches~\cite{hp_app_store}, and even drones~\cite{drone}. Yet the success that has been achieved through the crowd-sourcing app development on these platforms has raised concerns in both industry and academia about whether the autonomous vehicle (AV) -- the next much-anticipated software platform -- will become appified.  

\begin{figure}
\centering
\includegraphics[width=\columnwidth]{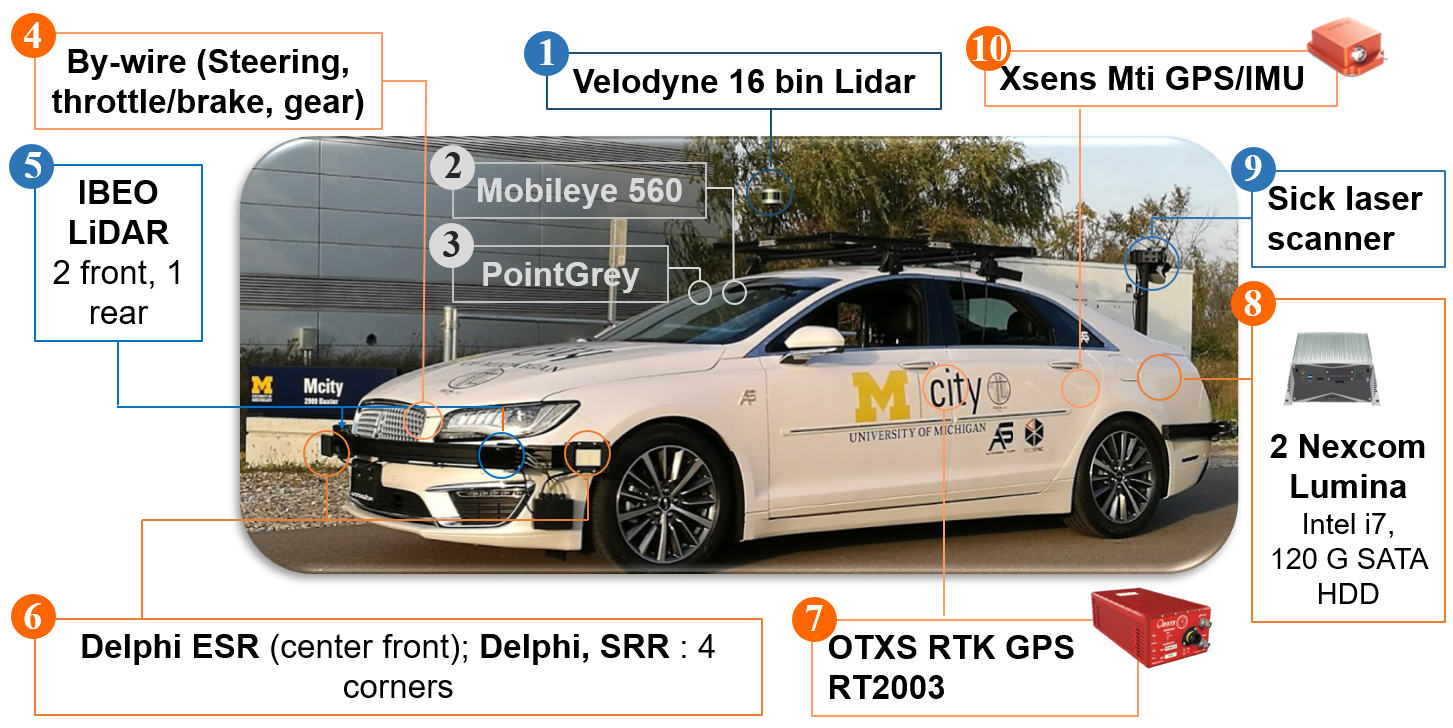}
\caption{The open-access automated vehicle of at the University of Michigan}
\label{fig:photo}
\end{figure}

Supporting crowd-sourcing app development on AV is presumed to benefit both automakers and customers. From the end-user's perspective, it will bring a vast variety of apps into the market that enriches users' choices, providing them with the flexibility to personalize their driving and in-vehicle experience just by installing/uninstalling apps. For the automakers, appification will transform the development of some AV functionalities from outsourcing to crowd-sourcing, which not only reduces the cost, but also promotes the improvement of app quality. The AV app market will also create an ecosystem where multiple functionalities can work together to provide greater intelligence and convenience. Some organizations have already begun to roll out the autonomous vehicle with open development support. For example, in the industry, vehicle middleware platforms that open massive vehicle functionalities, including steering wheel and brake to the developers have been built (e.g., Ford OpenXC~\cite{openxc}, PolySync~\cite{polysync}). While in the academia, University of Michigan (U-M) starts to offer open-access to their testing AV equipped with sensors, including lidar, radar, and cameras, so that researchers can rapidly test their self-driving or connected-vehicle technologies~\cite{umich_news} (Fig.~\ref{fig:photo}). 
Although these initiators have yet to discuss the idea of ``AV app store'', to better realize the benefit of crowd-sourcing development, there are reasons to believe that appified an AV platform will become a reality in the near future.



\begin{figure}
\centering
\includegraphics[width=\columnwidth]{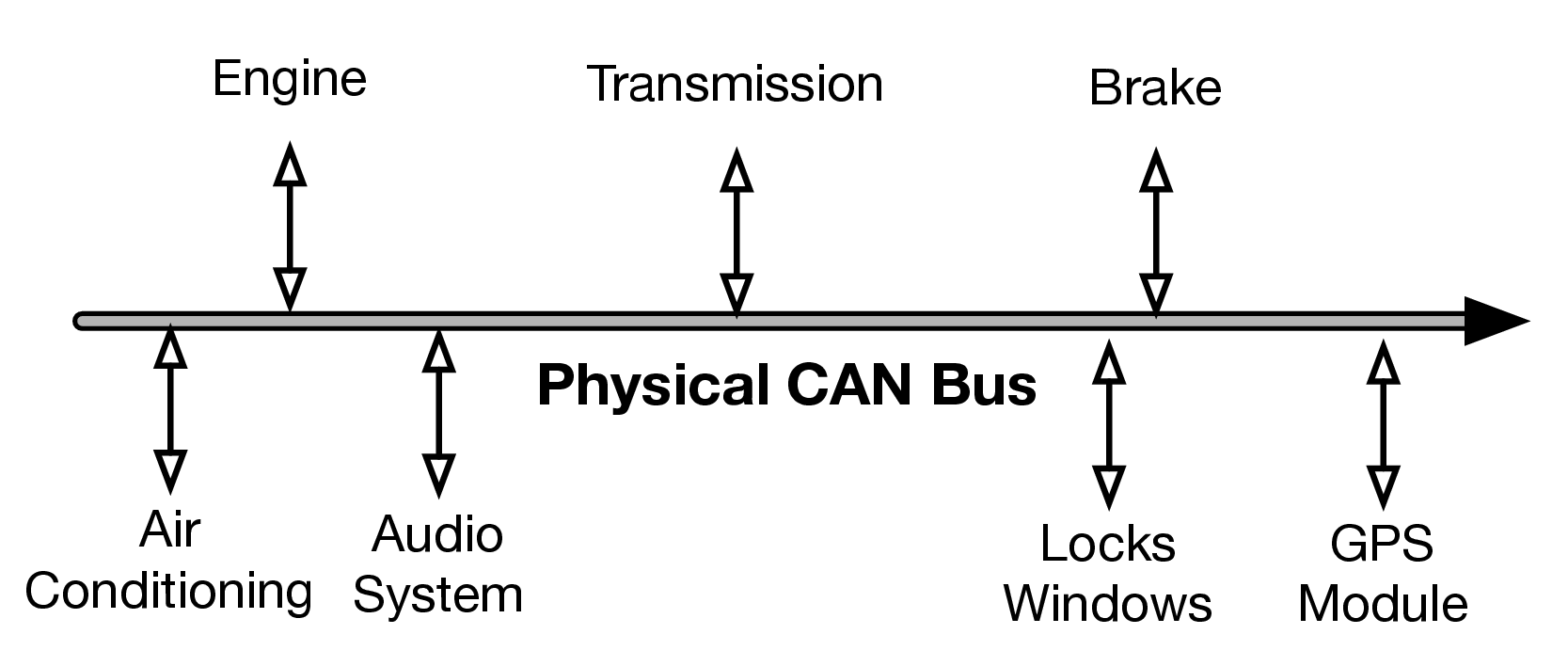}
\caption{Traditional physical CAN bus vehicle platform}
\label{fig:traditional}
\end{figure}

\begin{figure}
\centering
\includegraphics[width=\columnwidth]{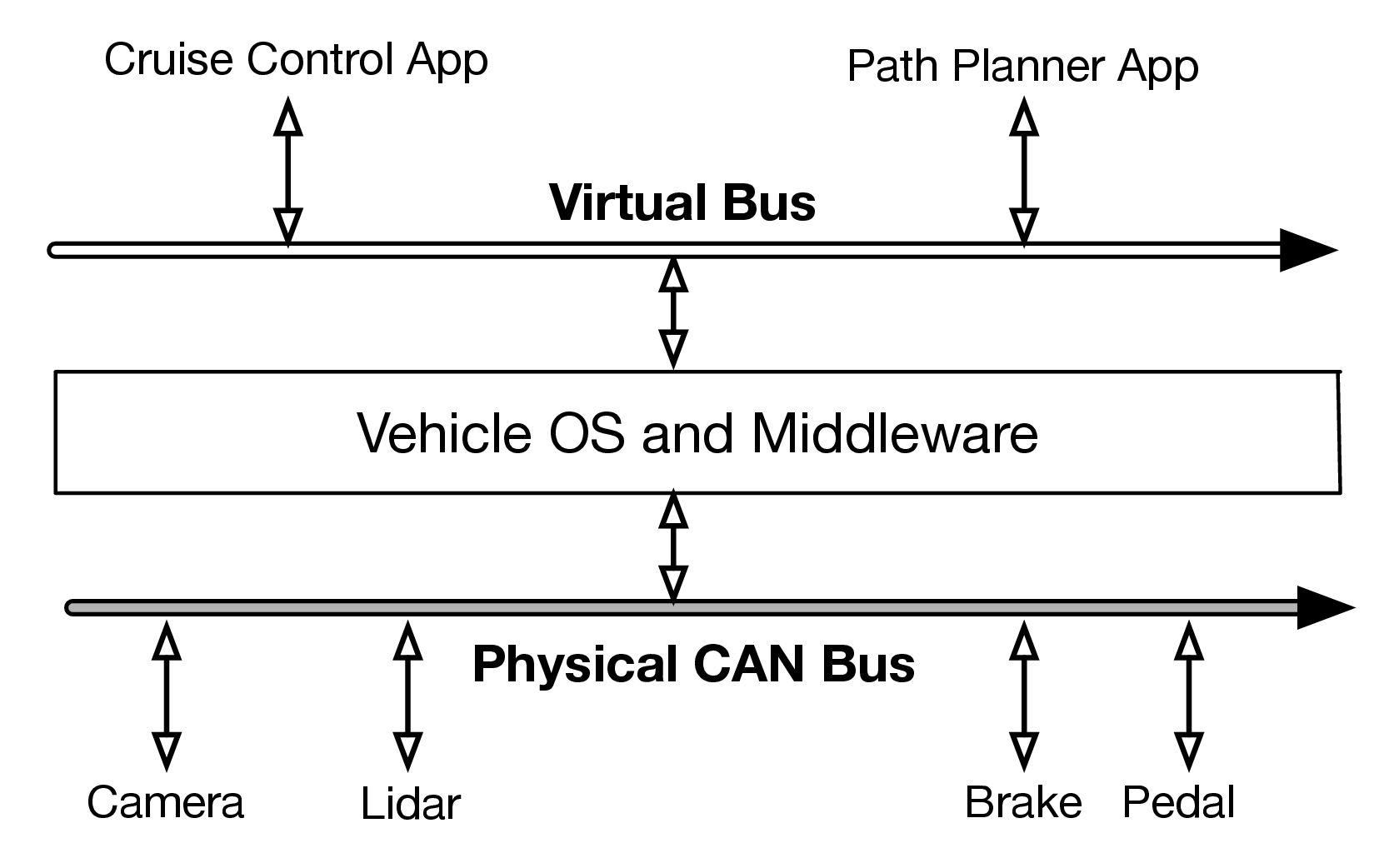}
\caption{Example of the appified AV architecture}
\label{fig:appified}
\end{figure}

There are two sides to every coin, however. To identify key challenges and issues for the appified AV, we compare aspects of the traditional vehicle platform with an appified AV platform. As can be seen in Fig.~\ref{fig:traditional}, from the adoption of the Controller Area Network (CAN) bus in the 1980s, modern vehicles now have over 70 electronic control units (ECUs) for various subsystems~\cite{albert2004comparison}, including critical control systems and also peripheral infotainment systems all communicate using the CAN bus. As automakers usually outsource the development of these peripheral functionalities to reduce development costs, security and safety problems arise,as software developed by a third-party with access to CAN bus, can potentially be exploited to tamper the safety of the vehicle~\cite{jeep_hack,mazloom16security}. The appified AV platform shown in Fig.~\ref{fig:appified} provides software abstraction for the physical CAN bus. Self-driving functionalities are developed as apps, and their interactions with the hardware actuators are proxied by a vehicle operating system that also acts as the middleware. Since the appified platform opens full-fledged self-driving functionalities to the third party developers, it could potentially introduce greater safety risk. For example, flaws in the proportional control algorithms of a cruise control app may put the vehicle in a situation where a collision becomes inevitable. Recent accident records~\cite{tesla_accident} of self-driving cars suggest that deficiencies in the AV software are inevitable due to the complexity of the physical environment, thus rendering the vulnerable apps a persistent threat to the AV industry. Apps may also be developed for malicious purposes to tamper with the user's safety with embedded malicious logic~\cite{car_malware}.  



Fortunately, the open nature of the appified AV platform also provides us with opportunities to build and deploy defenses against these potential threats at the vehicle OS level. Thus, sanity checks can be performed on the control messages from AV apps to mitigate potential risks raised by apps and guard the safety of the vehicle. However, due to the extremely high safety requirement of the vehicle platform compared with other appified platforms, dangerous apps that include fatal design and implementation flaws or malicious logic should really be detected even before being installed on the vehicle. The best practices that have already achieved success on appified platforms are building market-level app vetting to prevent problematic apps from entering the market. We draw lessons from state-of-the-art security practices in the IT field, and leverage the open AV platform to propose our solution to the above mentioned problem.


In this paper, we propose the first appified AV design scheme that focuses mainly on addressing security and safety concerns raised by crowd-sourcing app development. 
We identify key principles that need to be enforced in the context of various self-driving scenarios, and propose \name, which is a platform enhancement for open AV that incorporates (1) offline app vetting which performs early detection of unsafe apps, and (2) runtime safe guardian which provides a baseline safety guarantee. We present the detailed design of the app vetting process, which adapts the program analysis techniques from cybersecurity fields to perform static app analysis, and leverages recent advancements in the Accelerated Evaluation field to dynamically estimate the risks of apps using naturalistic traces. The vetting also enforces many principles proposed recently in the U.S. Department of Transportation's Federal Automated Vehicles Policy~\cite{dot_policy}, such as requirements for the fall-back approach and the specification of allowable circumstances. We also advocate a runtime on-vehicle watchdog implemented as a privileged process on the vehicle OS that monitors the surrounding environment to avoid potential collisions caused by apps. Our study sheds light on the opportunity of realizing the benefits of crowd-sourcing development on self-driving vehicles without risking the safety of the platform, and provides guidelines for future research along this line. 



%% file: secs/overview.tex
\section{The \name Approach}
\label{sec:overview}

In this section, we first describe the scope of the problem , and discuss the essential steps to mitigate them. We then introduce the major components of the \name approach, and present a roadmap for the detailed design of each component. 

As mentioned earlier, the major safety issue with appified AV comes from the untrusted third-party app. In this paper, we focus mainly on two types of untrusted apps -- \emph{vulnerable} app and \emph{malicious} app, which have been consistently presented difficulties for other appified platforms. A vulnerable app is one that contains flaws that may either cause damage directly to vehicle functionality or expose vulnerabilities that can be leveraged by an adversary. In addition to the complexity of self-driving functionalities, these mistakes may also be made due to inexperienced or careless developers. In contrast, a malicious app (malware) is one that intentionally tampers with the user's safety or privacy with malicious logic embedded in the apps. For example, malware may be disguised as a normal AV app, and block certain critical functionalities while the user is driving unless a ransom is paid. In some cases, the erroneous or malicious logic can be identified offline by analyzing the app's code, thus preventing suspicious apps from entering the market. The problem of some apps, however, is that they may be revealed only in certain roadside or environmental conditions, which would not be able to be identified statically. 

\begin{figure*}
\centering
\includegraphics[width=\textwidth]{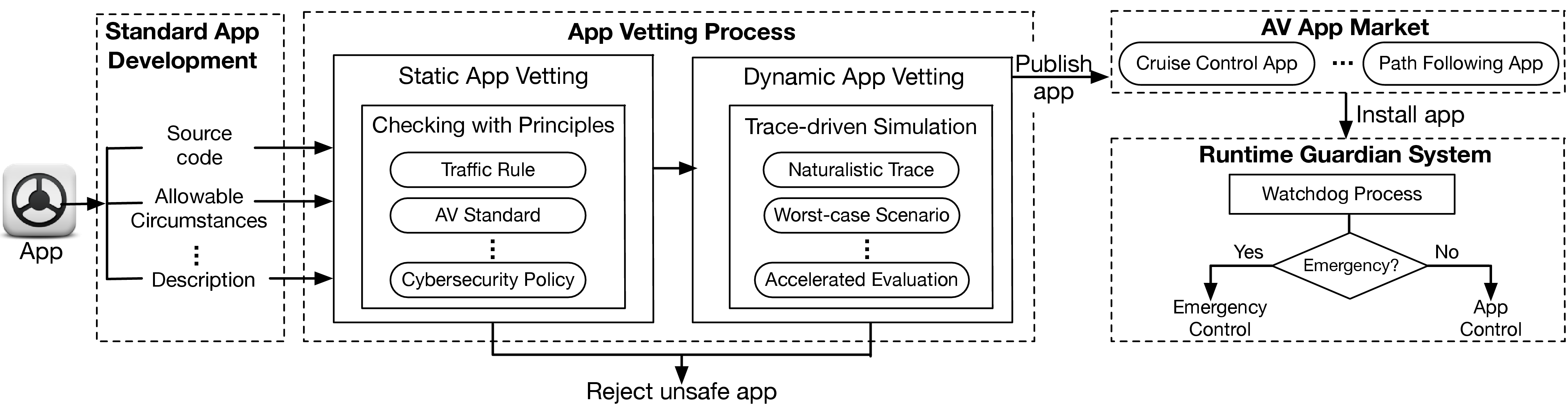}
\caption{System overview}
\label{fig:architecture}
\end{figure*}


As shown in Figure~\ref{fig:architecture}, we advocate the \name approach, which consists of four components: (1) A standardized app development process to be provided by the developer that specifies required information on the properties of the self-driving app (e.g., source code, allowable circumstances), in order to make the app functionality expressive and verifiable (\S\ref{sec:standard}); (2) A static app vetting  framework that checks app logic against a set of safety principles based on static program analysis techniques (\S\ref{sec:static}); (3) A dynamic risk evaluation system that tests the app in a simulated environment against a set of benchmarks and naturalistic driving traces to quantify potential risk (\S\ref{sec:dynamic}), with only apps that pass both static and dynamic vetting being allowed to enter the market. The official app market should be the only authenticated source from which users should be allowed to download AV apps. Authenticity can be enforced by requiring the installation of a digital signature for each app, with only app binaries signed by the official market being allowed to run on the consumer's vehicle; and (4) A runtime guardian system that performs access control for AV apps running on the vehicle based on the environmental context, and ensures the baseline safety of the vehicle under a variety of physical scenarios (\S\ref{sec:guardian}). We then discuss details about of implementation of \name in \S\ref{sec:discussion}, and conclude our work in \S\ref{sec:conclusion}.

%% file: secs/standard.tex
\section{Standardized AV App Development}
\label{sec:standard}

%
%
Existing appified platforms require app developers to provide a specification or statement about the security sensitive capabilities app demands\cite{apk_structure}. For example, in the appified smartphone platform with dominating market share, Android, each app's code base is required to provide a manifest file that lists the requested capabilities, such as reading and sending text messages. Such explicit specifications have already been shown in previous work in both academia and industry to greatly benefit the app vetting process, that is, by efficiently identifying malicious apps that masquerade as benign apps on the app market~\cite{enck2011study,app_vetting}.

Applying from the design of existing appified platforms to \name, we require that the developers of AV apps explicitly provide essential information to facilitate the app vetting process (detailed later in~\S\ref{sec:static} and~\S\ref{sec:dynamic}) and determine that the app behaviors conform to user expectations. Adapting previous design to the context of autonomous vehicle control, our preliminary design of the app specifications in \name includes three types of information: \textit{required resources}, \textit{purpose statement}, and \textit{usage constraints}. \textit{Required resources} are entries that specify the security or safety sensitive of in-vehicle resources the app needs to access. Examples of such resources include sensing devices such as radar, controlling devices such as wheels and throttle pedals, and also platform level resources such as networking, storage.

In addition to the list of resources requested, we also require that the developers specify why these resources are requested in the \textit{purpose statement} entries. In AV systems, the granting of access to certain resources is ultimately an end user decision, thus providing more detailed explanations for access to sensitive resources which can greatly help users make better decisions. In our preliminary design, there are two types of purpose statement entries: (1) App-level purpose statement answering the question as to what the app is used for, which makes it easier for users to choose the desired app; and (2) Resource-level purpose statement clarifying why certain resources (e.g., control of brake, speed of vehicle) are required, which can help permission -- granting decisions and also enable early detection of potential resource usage conflicts in the vetting process. 

\begin{lstlisting}[caption=Code snippet of a working path following app, language=C++, label=listing:example, float=tp]
class PathFollowingNode
{
	//App is a node on Virtual CAN bus
	List<Position> map;
	public PathFollowingNode(){
    //Subscribe on the vehicle report 
      registerListener( VEHICLE );
    //Load the trajectory data 
      map = loadMap("map.dat");
	}
	//When vehicle report message is received
	void messageEvent (Message msg)
	{
		float _distErr;
		float _headingErr;

		for (long i=startNum; i<map.length(); i++)
		{	//Calculate distance error and heading error referring to the path
			_distErr = sqrt(pow(msg.position.x-map[i].x,2)+
				pow(msg.position.y-map[i].y, 2));
			_headingErr = abs(msg.position.heading-map[i].heading);

			Message message = new Message();
			if(_distErr<MAX_DIST_ERR && _headingErr<MAX_HEADING_ERR){
			//If vehicle in path, continue with the adjusted steering angle	
				float _angel = CalcSteeringAngel(msg);
				message.setSteeringAngel(_angel);
			//Publish message on virtual CAN
				message.publish();
			}
			else{
			//If vehicle not in path, perform stop with the maximum throttle gain
				message.setThrottleCommand(100);
				message.publish()
				break;
			}
		}
	}
}
\end{lstlisting}

In \name, developers are also required to specify \textit{usage constraints} for each resource usage request. Due to the uncertainty of real-world road conditions, the apps in AV systems are usually designed for special/strict usage conditions, e.g., highway only, sunny day only (with clear camera vision). Thus, to provide safety guarantees, each app in \name must specify \textit{allowable circumstances} for the usage of the app itself. This requirement is unique for the autonomous vehicle domain, and is also listed in the DoT's federal automated vehicle policy. In addition to the app-level allowable circumstances, we also ask developers to specify resource level usage constraints. For example, a cruise control app may need to specify an exclusive usage requirement of the steering wheel to ensure functionality correctness.

Listing~\ref{listing:example} is the code snippet of a working self-driving app developed by a U-M researcher that implements the path following functionality for the AV and enables the vehicle to drive following a given set of coordinates. We use it as a running example to help understand our approach in the rest of the paper. For this app, the required resource part of the app specification needs to include the prerequisite to subscribe to the vehicle state message, in order to access vehicle speed and heading angle. It also needs to specify its demand to control the steering wheel and the throttle. The purpose statement can explain that the app-level purpose is to provide trajectory following and the resource-level purpose is to monitor the vehicle state update, and to perform the  steering wheel angle adjustment and emergency stop for the steering and throttle access. Since this app is designed to follow only a given path, its app-level usage constraint is to be used only in the proprietary test facility (e.g., Mcity~\cite{mcity}), where traffic is under control, and its resource-level constraint can be having exclusive use of the steering wheel and throttle access.

In our design, these specifications provided by developers must required to be written in a certain language or format to facilitate automated processing. Extensible Markup Language (XML) is one recommended format that can be used to write specifications that are both human-readable and machine-readable~\cite{xml}. The standardization not only helps the app vetting process to identify malicious or vulnerable apps, but also enables early detection of app-level and resource-level conflicts. It will also make it easier for users to choose the desired app. 



%% file: secs/static.tex
\section{Static App Vetting}
\label{sec:static}

Static program analysis is a software engineering method that leverages standard a program language techniques to automatically examine the source code of a program to check a set of pre-defined code properties~\cite{wogerer2005survey}. In existing appified platforms, this technique has already been demonstrated to be effective in checking a wide range of security related properties, including malicious app behaviors such as malicious code execution and information leakage, and also app vulnerabilities such as vulnerable API usage~\cite{chen2015static,arzt2014flowdroid,wei2014amandroid}. For example, static analysis on AV apps is able to identify a potentially malicious infotainment app if it contains logic for publishing a steering wheel control message in the code.

\begin{figure}
\centering
\includegraphics[width=\columnwidth]{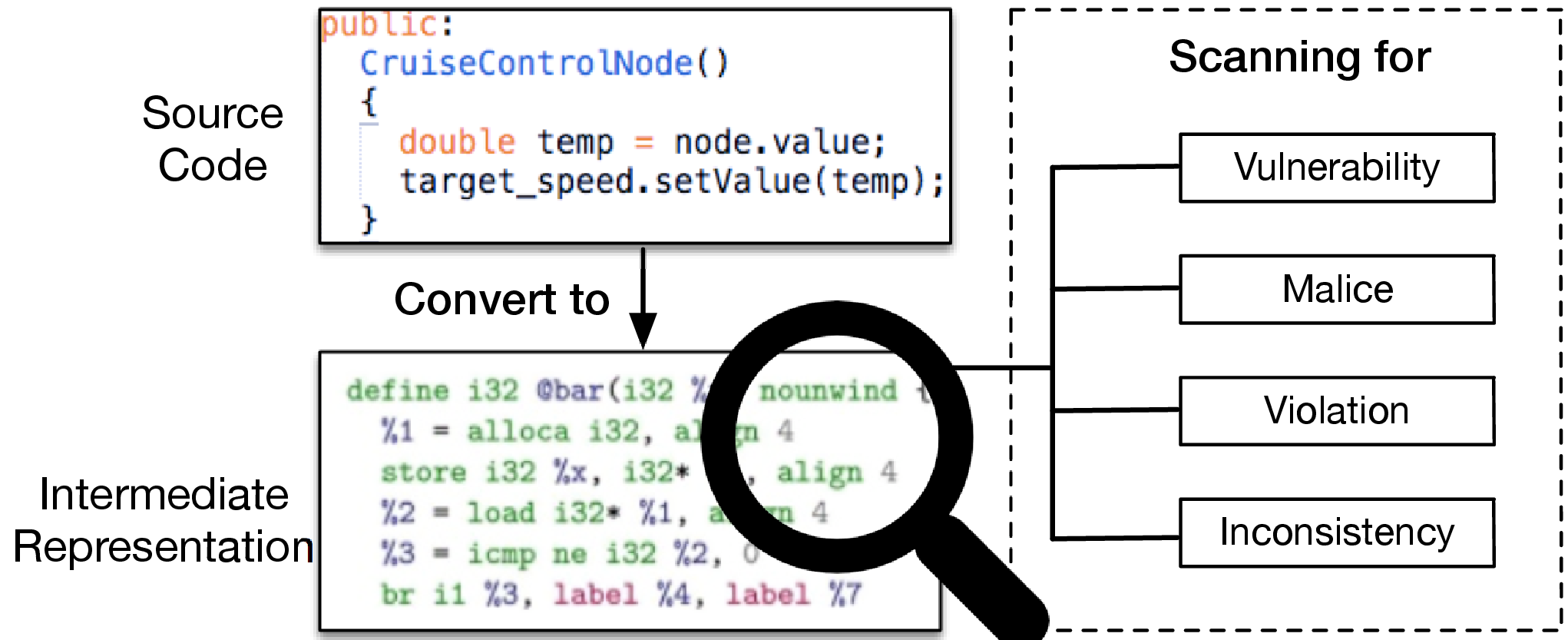}
\caption{Static analysis overview}
\label{fig:static}
\end{figure}

In \name, we use static analysis as the first step in our vetting process. Fig.~\ref{sec:static} shows an overview of the analysis procedure. It first converts the AV app's source code written in a particular programming language to an Intermediate Representation (IR)~\cite{ir} that can be automatically processed by state-of-the-art programming analysis approaches. The code properties our system is designed to check for include (1) known security sensitive properties, e.g., the buffer overflow vulnerability in C programs; (2) known malicious code patterns such as collecting user privacy data (e.g., location) and sending it to suspicious hosts, (3) consistency with the specifications submitted by the developer together with the source code ; and (4) safety sensitive properties specific to the AV platform. For the latter, our system derives such properties from multiple sources, including, but not limited to:
\begin{itemize}
\item \textbf{Existing rules applied to commodity vehicle components.} Protocols such as MirrorLink~\cite{mirrorlink} have already brought the concept of isolation to the infotainment module of the vehicle, and these rules will also be enforced on the appified platform. 
\item \textbf{Traffic rules.} Will by default be enforced by the platform, so that violations in the app logic, such as not reacting to a red light, can be detected. 
\item \textbf{General cybersecurity policies.} Protecting the privacy of the app user is one example of such cybersecurity policies. Unauthorized access and transmission of sensitive in-vehicle information, such as location and user's profile, will be disallowed.
\item \textbf{Other vehicle-specific rules.} Some principles that seem to be common sense for human drivers may be violated in the appified era. For example, whether the gear has been shifted to 'Park' before turning of the engine can be verified from the app code. 
\item \textbf{AV industry standard.} Although the AV industry is still immature, with few published regulations. We expect industry standards to be formalized in the near future, with every app written for AV being checked against these standards. 
\end{itemize} 

Checking these properties is greatly facilitated by the specifications provided by the app developer in the app package as described in \S~\ref{sec:standard}. The static analysis engine automatically processes the app's purpose, scope and required resources in the specification. Program techniques are then used to realize the app vetting process, e.g., checking the conformation of the app's declared purpose and the actual implemented logic for malicious app detection.


%% file: secs/dynamic.tex
\section{Dynamic App Vetting}
\label{sec:dynamic}

\begin{figure}[t]
  \centering
  \subfigure[Desired and measured trajectories]{\includegraphics[width=0.34\columnwidth]{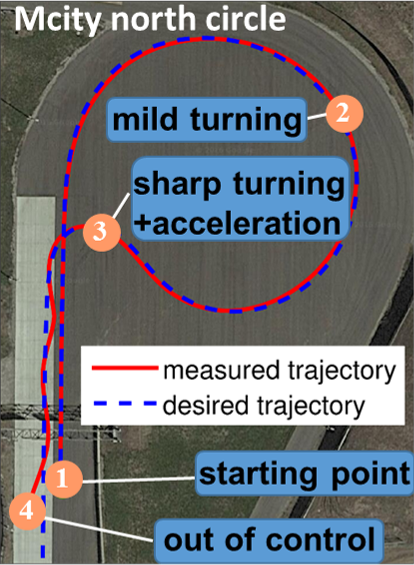}}
  \subfigure[vehicles states (longitudinal speed, yaw rate, steering angle)]{\includegraphics[width=0.64\columnwidth]{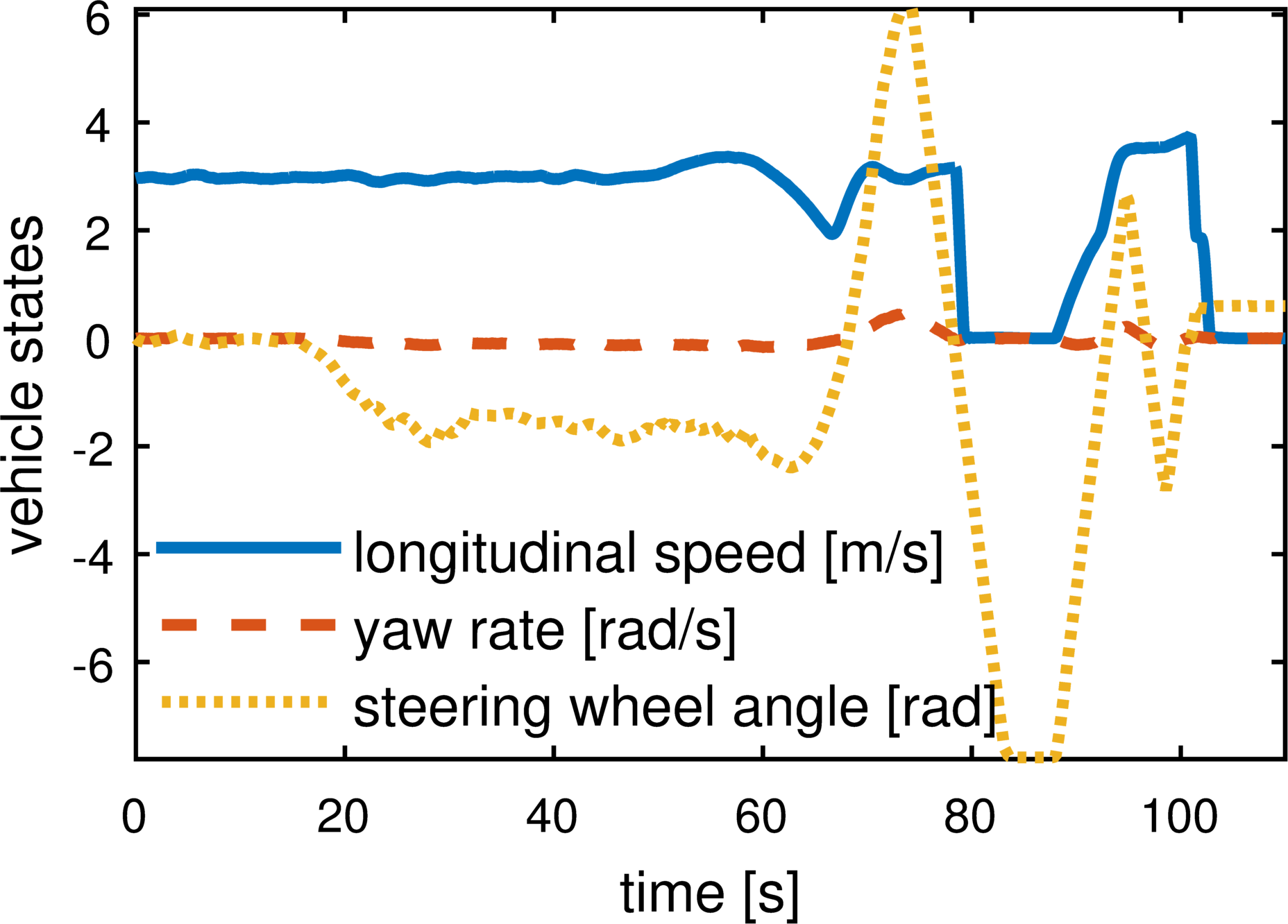}}
  \caption{Example of improper dynamic controller that led to out-of-control failure }
  \label{fig:Mcity_test}
\end{figure}

Static app vetting is effective for checking the app logic with invariant principles by ensuring perfect code coverage in the analysis. Some potential risks may not be revealed, however, until the vehicle encounters certain physical roadside conditions. 
Using the same path following app as an example (Listing~\ref{listing:example}), the vehicle status of running the app is plotted in Fig. \ref{fig:Mcity_test}. The automated vehicle successfully followed the trajectory when the acceleration and curvature were small (point 2), but lost control when the desired yaw rate was high (point 3). The vehicle completely lost stability at point 4  where the driver needed to take over to avoid a collision. Dynamic analysis bridges the gap by testing the program ideally in all the scenarios for which it is designed to detect any potential violation of security and safety principles that need to be upheld under any circumstance.

Dynamic analysis has achieved great success in detecting malicious and vulnerable apps on other appified platforms by eliminating the need to artificially create situations likely to produce errors. However, the known limitation is that it functions only as well as the principles used against which the code is being checked. Due to the complexity of the physical conditions in the self-driving scenarios compared, it virtually impossible to exhaustively list the safety principles suitable for individual scenarios~\cite{anderson2008use,shoshitaishvili2016sok}. We turn now to build a metric that can quantitatively evaluate the safety of a given AV app. 

\begin{figure*}[h]
\centering
\includegraphics[width=\textwidth]{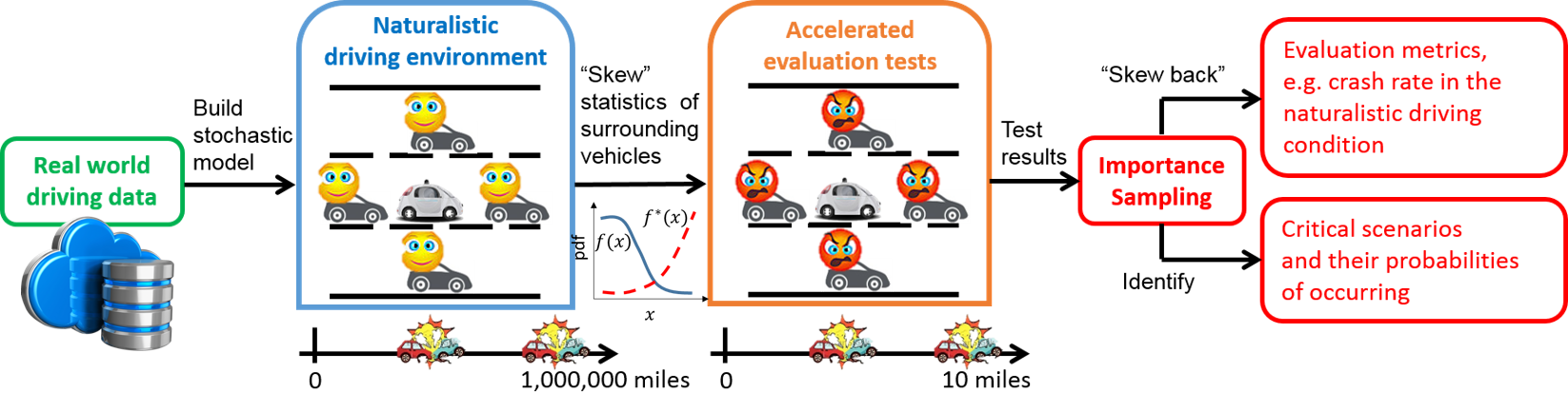}
\caption{Accelerated Evaluation overview}
\label{fig:AE}
\end{figure*}

Naturalistic Field Operational Tests (N-FOTs) have been used to evaluate AVs with data collected from a number of equipped vehicles driven in naturalistic conditions~\cite{festa2008festa}. Challenges that arise, however, when applying N-FOTs on an appified AV are: (1) it is too time consuming to thoroughly test each app, since critical conditions that likely lead to a crash are exceedingly rare in naturalistic traces, and (2) it still requires the app to be installed and tested on a real vehicle, which risks property and is not scalable for performing market-level vetting. We therefore advocate a simulation-based dynamic app vetting process that leverages a technique we call “Accelerated Evaluation” \cite{Zhao2016g,Zhao2016AcceleratedTechniques,Huang2016UsingScenario}, which consists of both the modeling aspect from data and the simulation algorithmic design. The procedure begins with calibrating the AV control dynamics and model-fitting the surrounding stochastic environment using naturalistic driving data from the naturalistic driving database. Once the distributions governing these stochastic elements are identified, the Importance Sampling method is applied, which entails finding a “skewed” distribution to artificially boost the hit rate of the rare event. Finding a good skewed distribution is conducted via adaptive searching method that sequentially optimizes the parameters in the candidate set of parametric distributions. (See Fig. \ref{fig:AE} for a pictorial outline of the procedure.)



The dynamic app vetting of \name is also performed off-the-vehicle, and aims at quantifying the potential risks of AV apps. Submitted apps whose probability of a crash in the trace-driven test exceed a certain threshold will be rejected. The simulation is performed by running the vehicle OS on a commodity server with naturalistic data feeding into the virtual CAN bus of the OS. Apps to be evaluated run on the ``simulated vehicle'' as if they were interacting with a real vehicle, read sensor data from the recorded trace, and send control messages to the virtual CAN. The vehicle OS simulates the app control on the current virtual environment, and adjusts the simulated sensor date to reflect the post-control environment. Thus, full-fledged self-driving functionalities of the app can be tested before installation on the real vehicle. As most of the existing vehicle OS are built based on the Robot Operating System (ROS), which runs on commodity computers, the effort of to build the simulation can be greatly reduced.

Accelerated Evaluation statistically models the driving behaviors reflected in the naturalistic trace, and compresses the non-safety-critical events to reduce the required test mileage for crash analysis by a factor of 10,000 to 100,000. Leveraging Accelerated Evaluation to statistically tailor the collected trace, the simulation can efficiently give the probability of a crash for a given app in the trace-driven simulation equivalent to an expanded driving test. Note that based on the allowable circumstances specified by the app, different sets of naturalistic trace will be used. For example, a cruise control app designed specifically for highway driving will be tested solely against highway traces extracted from the dataset. In addition to naturalistic traces, the compound approach combining static and dynamic vetting prevents threats of untrusted apps from entering the market.  

%% file: secs/guardian.tex
\section{On-Vehicle Runtime Watchdog}
\label{sec:guardian}

Even with static and dynamic app analysis, safety is not guaranteed under all circumstances, as chances still exist that some rare scenarios may be missing in the naturalistic trace. Moreover, interactions among different installed AV apps at runtime are also not covered by the analysis, and potential conflict in different app logic may lead to an accident. We maintain that an on-vehicle safety watchdog is essential for providing baseline safety guarantees. 

The proposed watchdog runs as a high-privileged process on the vehicle and can intercept AV apps' control messages and decline/override them. As shown in Fig.~\ref{fig:architecture}, the watchdog process senses the surroundings and the roadside conditions of the vehicle, and runs a basic collision avoidance algorithm. When it detects a potential collision caused by the self-driving app, it performs simple emergency reactions such as emergent stop or staying in lane to avoid collision; further commands issued by the apps will then be invalided. Note that different from other self-driving functionalities, the watchdog process provides a fail-safe mechanism only in unexpected roadside conditions that provides safety based on the sensing of the physical environment. The code base of the watchdog process implementation should be small, and the safety and reliability of the algorithm should be verifiable.


%% file: secs/discussion.tex
\section{Deployment Discussion}
\label{sec:discussion}
The platform we are targeting for prototyping the \name is a Lincoln MKZ equipped with PolySync middleware and a rich set of sensing and GPS devices including Mobileye and PointGrey cameras, Velodyne and IBEO Lidars, OTXS RTK GPS. To help other teams at the University test their control algorithms designed for self-driving cars on the road, we allow researchers to develop their own apps and test them on the vehicle.

\name satisfies the growing demands of opening the platform while preserving safety, even under the threat of potentially flawed or dangerous apps developed by students. We are also rolling out a standardized app development platform for all potential researchers interested in development on the AV which requires that the apps first be submitted in the source code format to our web-based vetting system. Only those apps that have passed the vetting will be compiled to executable binaries by the server and installed automatically on the real vehicle. The platform is also expected to be used in the upcoming UM course on Connected and Automated Vehicle in the fall of 2017 as an educational platform for students to safely test their self-driving algorithms on the road.

%% file: secs/conclusion.tex
\section{Conclusion}
\label{sec:conclusion}
We present the design of \name, which leverages state-of-the-art program analysis techniques and accelerated vehicle evaluation theory, integrating both to mitigate the security and safety challenges of running untrusted code on an open AV platform. Our approach is one solution to the challenge of realizing the functionality benefit of crowd-sourcing development in the arena of self-driving cars, without risking safety, and it is closely aligned to the increasing demand for self-driving car apps. We feel that \name provides guidelines for the future design of open AV, and as our future work, we will prototype it on our AV platform to enable the crowd-sourcing development in multiple scenarios. 